\documentclass{aa}  
\usepackage{orcidlink}
\usepackage{etoolbox}
\usepackage{graphicx}
\usepackage{txfonts}
\usepackage{xcolor}

\begin{document}

   \title{Simulation of Impact-induced seismic shaking on asteroid (25143) Itokawa to address its resurfacing process}

   \author{Sunho Jin
          \inst{1,2}
          \orcidlink{0000-0002-0460-7550}
          \and
          Masateru Ishiguro\inst{1,2}
          \orcidlink{0000-0002-7332-2479}
          }

   \institute{Department of Physics and Astronomy, Seoul National University, 1 Gwanak-ro, Gwanak-gu, Seoul 08826, Republic of Korea
   \and SNU Astronomy Research Center, Department of Physics and Astronomy, Seoul National University, 1 Gwanak-ro, Gwanak-gu, Seoul 08826, Republic of Korea\\
              \email{jsh854@snu.ac.kr, ishiguro@snu.ac.kr}
             }

   \date{Received; accepted }

% \abstract{}{}{}{}{} 
% 5 {} token are mandatory
 
  \abstract
{The surface of asteroid (25143) Itokawa shows both fresh and mature terrains, despite its short space weathering timescale of approximately $10^3$ years, as inferred from recent studies.
%despite its short space weathering timescale of approximately $10^3$ years \citep{Jin2022}. 
Seismic shaking triggered by the impact that formed the 8-meter Kamoi crater has been proposed as a possible explanation for the diversity.
}
{This study aims to examine whether the seismic shaking induced by the impact could account for the observed spatial variations in space weathering and further constrain the internal structure of Itokawa.}
{Assuming that the Kamoi crater was formed by a recent impact, we conducted three-dimensional seismic wave propagation simulations and applied a simplified landslide model to estimate surface accelerations and boulder displacements.
}
{Our results show that even a low-energy case (1\% of the nominal seismic energy) produces surface accelerations sufficient to destabilize the surface materials. The simulated boulder displacements are consistent with the observed distribution of space weathering degrees even on the opposite hemisphere. We estimate the seismic diffusivity to be 1\,000-2\,000 $\mathrm{m^2\,s^{-1}}$ and the seismic efficiency to be in the range of $5.0 \times 10^{-8}$ to $5.0 \times 10^{-7}$, implying that Itokawa’s interior contains blocks tens of meters across and acts as a strongly scattering medium.
}
{Our findings provide unique dynamical evidence, based on seismic wave propagation modeling, that supports the hypothesis of a truly rubble-pile interior for Itokawa.
}

               \titlerunning{Seismic shaking on asteroid (25143) Itokawa}

   \maketitle
%
%________________________________________________________________
\section{Introduction} \label{sec:intro}
The surfaces of airless bodies undergo various physical and chemical evolutionary processes when exposed to the space environment. Among these processes, space weathering, caused primarily by solar wind implantation and micrometeorite bombardment, is one of the most significant processes altering their optical properties \citep{Pieters2016, Chapman2004, Clark2002}. This process has been observed on several celestial bodies, including the Moon \citep{Pieters2000}, Mercury \citep{Hapke2001}, and many asteroids \citep{Clark2002}. In particular, for S-type near-Earth asteroids, space weathering is known to reduce albedo, redden the spectral slope, and reduce the absorption depth near 1 $\mu$m \citep[see e.g.][]{Sasaki2001}.

The Hayabusa mission target, asteroid (25143) Itokawa, exhibits a unique appearance characterized by a wide range of space weathering degrees across its surface. However, a puzzling discrepancy exists in the estimated timescales of weathering on Itokawa, ranging from $10^2$ to $10^7$ years \citep{Bonal2015,Koga2018,Noguchi2011,Keller2014,Matsumoto2018,Nagao2011}. Based on analyses of returned samples, space weathering research has made significant progress. \citet{Nagao2011} estimated surface exposure ages of only 3\,000 to 8\,000 years through noble gas analyses of returned particles. In addition, \citet{Noguchi2014} observed nanometer-scale weathered rims on the surfaces of Itokawa grains using transmission electron microscopy, offering new insights into the microstructural development of space weathering. Our group also estimated a space weathering timescale of approximately $10^3$ years by analyzing the size-frequency distribution of white mottles on boulder surfaces \citep{Jin2022}. This result, based on remote-sensing data, is consistent with  estimates from returned samples by \citet{Noguchi2011}, which indicate short exposure durations.

Despite the short timescale of space weathering on Itokawa, it remains unclear why several regions still display less weathered surfaces. If the impact had occurred more than several tens of thousands of years ago, such fresh features would likely have been erased by space weathering. In addition, fresh terrains are not only found near the Kamoi crater but are also scattered across other regions, and the factors controlling this spatial distribution remain unresolved. \citet{Saito2006} noted that fresh terrains are more prevalent on the Kamoi crater side, but did not address the cause of their formation. \citet{Jin2022} proposed that the 8-meter-wide Kamoi crater may have generated seismic waves strong enough to trigger localized landslides, thereby exposing unweathered material beneath the surface. They further hypothesized that areas with large boulders may display fresh terrains due to lateral boulder movement, which could uncover pristine material underneath.

To test this hypothesis proposed in \citet{Jin2022}, we conducted three-dimensional (3D) seismic wave propagation simulations and subsequent landslides resulting from the Kamoi impact. Previous studies have modeled seismic wave propagation within asteroids (e.g., \citealt{Richardson2005, Quillen2019, 2024PSJ.....5..251B}). Notably, \citet{Richardson2020} estimated crater retention ages and constrained surface and internal properties of four asteroids, including Itokawa, by modeling topographic diffusion caused by seismic shaking. However, their study employed simplified cylindrical coordinates scaled to match the surface area of each asteroid, as part of a generalized approach, which limited the fidelity of seismic wave propagation analysis in a realistic shape model.

We utilized a shape model based on remote-sensing observations of Itokawa, reconstructed from disk-resolved images taken by the Hayabusa/AMICA instrument \citep{Gaskell2020}, to compute seismic wave propagation within the asteroid's three-dimensional geometry. In addition, we developed a toy model to quantitatively simulate landslides on each facet of the shape model. Based on the simulation results, we analyzed the conditions under which fresh terrains are formed on Itokawa. Furthermore, we investigated how varying physical parameters influence our results, particularly in terms of inferring the internal structure from the spatial distribution of surface space weathering.

\section{Method} \label{sec:method}

We employed a two-step modeling approach to investigate the spatial distribution of fresh terrains on the surface of Itokawa. First, we simulated the propagation of seismic energy triggered by the impact that formed the Kamoi crater using a three-dimensional diffusion model (Section \ref{subsec:propagation}). Then, we developed a simplified landslide model to estimate the displacement of surface boulders (Sect. \ref{subsec:landslide}). The following subsections describe each step in detail.

\subsection{Seismic energy propagation} \label{subsec:propagation}

\begin{figure*}
\centering
 \includegraphics[width=\hsize]{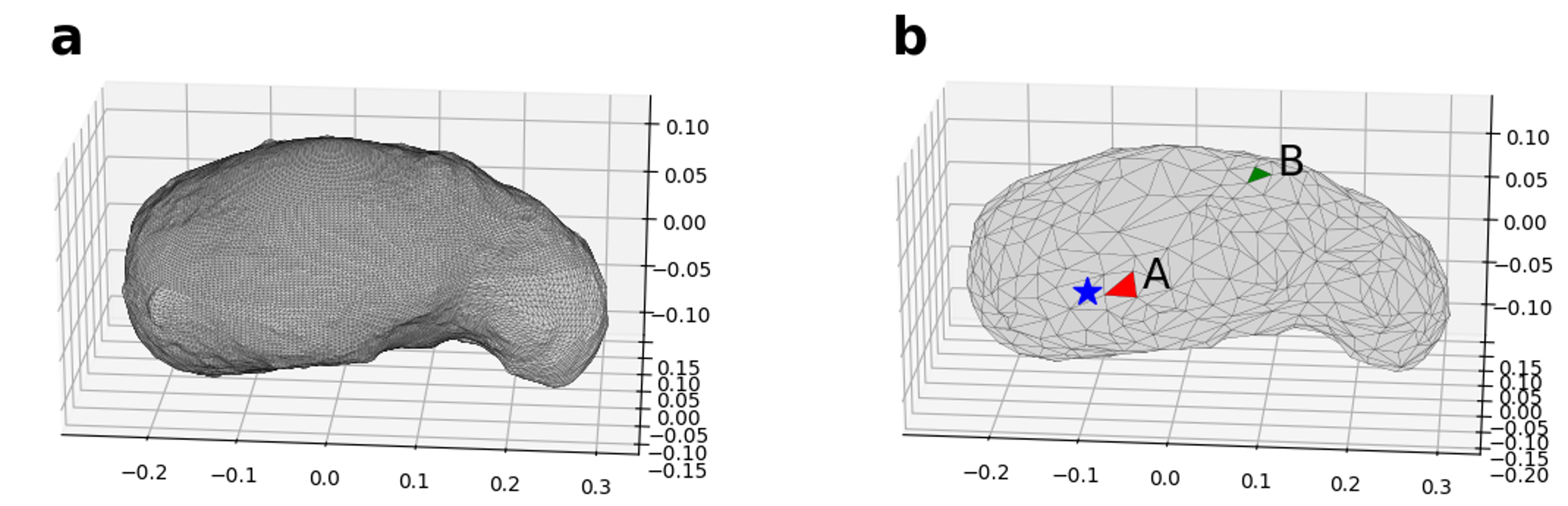}
 \caption{
3D views of (a) the original shape model of Itokawa with 49\,152 facets from \citet{Gaskell2020}, and (b) the simplified model with 1\,024 facets used in this study. The blue star denotes the location of the Kamoi crater. Red (A) and green (B) facets correspond to facet numbers 152 and 93, respectively, and are used for comparison in Fig.~\ref{fig:acceleration} and Fig.~\ref{fig:x_facet} }
\label{fig:simplify}
\end{figure*}

We conducted a numerical simulation for the seismic energy propagation using the shape model provided by \citet{Gaskell2020}, which contains 49\,152 triangular facets in its simplest form (Fig. \ref{fig:simplify}a). 
We found that some facet slopes in the model were affected by meter-scale surface boulders and do not reflect the real slope of the asteroid surface. 
Therefore, we reduced the number of facets to 1\,024 using \texttt{pymesh.meshing\_decimation\_quadric\_edge\_collapse}, a function from PyMesh, a Python library for geometry processing (see Figure~\ref{fig:simplify}b). This function is based on the quadric error metrics algorithm \citep{Garland1997}. As a result of the simplification, the average area of the simplified facets is $3.87 \pm 2.09 \times 10^2~\mathrm{m}^2$, which corresponds to an effective length of $19.03 \pm 4.94$ m. This length is larger than most surface boulders, effectively minimizing the influence of individual boulders on the facet geometry \citep{Michikami2008}.

Next, we prepared a 3D grid to solve the seismic energy diffusion equation. 
We generated a $600 \times 400 \times 300$ grid with 2 m spacing. 
The grid size and resolution were chosen so that the entire volume of Itokawa, whose maximum extent is approximately 535~m, is fully enclosed within the grid. 
We assigned an initial value of 0 to grid cells inside the shape model. 
Grid cells outside the shape model were assigned a placeholder value of NaN (Not a Number), a standard marker in numerical computations used to indicate undefined or excluded regions.

The total seismic energy $E_\mathrm{s}$ was calculated using the following equation:

\begin{equation}
    E_\mathrm{s} = \eta E_\mathrm{i} = \eta \frac{1}{2} m_\mathrm{i} v_\mathrm{i}^2 
    = \eta \frac{2}{3} \pi r_\mathrm{i}^3 \rho_\mathrm{i} v_\mathrm{i}^2 , 
    \label{eq:totalE}
\end{equation}

\noindent
where $E_\mathrm{i}$, $m_\mathrm{i}$, $v_\mathrm{i}$, $r_\mathrm{i}$, and $\rho_\mathrm{i}$ are the kinetic energy, mass, velocity, radius, and mass density of the impactor, respectively. 
We assumed $v_\mathrm{i} = 25~\mathrm{km\,s^{-1}}$, $r_\mathrm{i} = 20$ cm, and $\rho_\mathrm{i} = 3~\mathrm{g\,cm^{-3}}$ to reproduce an 8-m crater like Kamoi, based on the scaling relation by \citet{Tatsumi2018}. $\eta$ is an efficiency parameter, defined as the ratio between total seismic energy and the kinetic energy of the impactor. 
The values of $\eta$ used in the simulations are listed in Table \ref{table:sims}. 
\citet{Richardson2020} estimated the seismic efficiency of Itokawa to be $1.0 \pm 0.5 \times 10^{-7}$.

Using the Small Body Mapping Tool \citep{Ernst2018}, we identified the location of the Kamoi crater as $(X, Y, Z) = (-67.4, -143, 15.2)$. 
We assumed that the initial seismic energy was distributed radially within 4 m from this point, weighted inversely with the squared distance. This approach provides a simple approximation of localized energy deposition near the impact site. Although not based on a detailed seismic source model, the inverse-distance method concentrates energy near the impact and alleviates edge discontinuities. 

\begin{table}
\caption{\label{table:sims} List of simulations introduced in this work}
\centering
\begin{tabular}{cccc}
\hline \hline
No. \tablefootmark{a} & $\eta$  \tablefootmark{b} & $K~(\mathrm{m^2s^{-1}})$ \tablefootmark{c} & Note \\
\hline
1  & $1.0 \times 10^{-7}$ & $2\,000$     & \tablefootmark{d} \\
2  & $1.0 \times 10^{-7}$ & $1\,000$     & \tablefootmark{e} \\
3  & $1.0 \times 10^{-7}$ & $3\,000$     & \tablefootmark{e} \\
4  & $5.0 \times 10^{-8}$ & $1\,000$     & \tablefootmark{e} \\
5  & $5.0 \times 10^{-8}$ & $2\,000$     & \tablefootmark{e} \\
6  & $5.0 \times 10^{-8}$ & $3\,000$     & \tablefootmark{e} \\
7  & $5.0 \times 10^{-7}$ & $1\,000$     & \tablefootmark{e} \\
8  & $5.0 \times 10^{-7}$ & $2\,000$     & \tablefootmark{e} \\
9  & $5.0 \times 10^{-7}$ & $3\,000$     & \tablefootmark{e} \\
10 & $1.0 \times 10^{-9}$ & $2\,000$     & \tablefootmark{f} \\
11 & $1.0 \times 10^{-5}$ & $2\,000$     & \tablefootmark{f} \\
12 & $1.0 \times 10^{-6}$ & $2\,000$     & \tablefootmark{f} \\
13 & $1.0 \times 10^{-7}$ & $200$       & \tablefootmark{g} \\
14 & $1.0 \times 10^{-7}$ & $200\,000$   & \tablefootmark{g} \\
\hline
\end{tabular}
\tablefoot{
\tablefoottext{a}{Simulation number} \\
\tablefoottext{b}{Seismic energy efficiency} \\
\tablefoottext{c}{Diffusivity} \\
\tablefoottext{d}{Nominal case from \citet{Richardson2020}} \\
\tablefoottext{e}{Values within the range suggested by \citet{Richardson2020}} \\
\tablefoottext{f}{Extremely low or high $\eta$ cases} \\
\tablefoottext{g}{Extremely low or high $K$ cases} \\
}
\end{table}

%\begin{figure}
%\centering
%\includegraphics[bb=10 20 100 300,width=3cm,clip]{IS_initial.pdf}
%\caption{The initial seismic energy density ($\epsilon$) distribution for simulation 1 and others with the same $\eta$ values. Colors of each cube shows energy density within each grid.} \label{fig:initial}
%\end{figure}

We modeled the propagation of seismic energy using a diffusion equation with an attenuation term, following the approach of \citet{Dainty1974} and \citet{Richardson2005}:

\begin{equation}
    \frac{d \epsilon}{dt}  = K \nabla^2 \epsilon - \frac{2\pi f \epsilon}{Q},
    \label{eq:diffusion}
\end{equation}

\noindent where $\epsilon$ is the seismic energy density, $K$ is the diffusivity, $f$ is the dominant frequency of the seismic wave, and $Q$ is the quality factor. The diffusivity is defined as:

\begin{equation}
    K = \frac{1}{3} v l,
    \label{eq:diffusivity}
\end{equation}

\noindent where $v$ is the seismic wave velocity and $l$ is the mean free path. The values of $K$ used in the simulations are listed in Table~\ref{table:sims}. We adopted a nominal and possible range of diffusivity in \citet{Richardson2020}, where the authors reported $K = 2\,000~\mathrm{m^2\,s^{-1}}$ as a nominal value and $1\,000-3\,000~\mathrm{m^2\,s^{-1}}$ as a possible range. We assumed a dominant frequency of $f = 10$ Hz, consistent with hydrocode simulations showing that seismic wave energy peaks in this frequency range \citep{Richardson2020}. We adopted a constant quality factor of $Q = 1\,500$ for all cases.

We solved Eq. \ref{eq:diffusion} using the 4th-order Runge-Kutta method for 10 seconds after the impact. The default time step was set to 0.001 seconds, and reduced to 0.0001 seconds for cases where $K > 3\,000~\mathrm{m^2\,s^{-1}}$ to maintain numerical stability. We applied Neumann boundary conditions (zero energy flux) to model waves reflected from the surface since the outside of the asteroid has a much lower density than the interior. However, this assumption likely overestimates retained energy by ignoring losses to space. Such an effect of the boundary condition is discussed in Sect. \ref{subsec:uncertainty}. 
At each time step, we converted the seismic energy density $\epsilon$ into peak acceleration amplitude $a$, assuming simple harmonic oscillation:

\begin{equation}
    a  = \sqrt{\frac{8\pi^2 f^2 \epsilon}{\rho_\mathrm{t}}}~,
    \label{eq:acceleration}
\end{equation}

\noindent where $\rho_\mathrm{t} = 1.9~\mathrm{g\,cm^{-3}}$ is the bulk density of Itokawa \citep{Fujiwara2006}. 
This provides the acceleration envelope for monochromatic oscillation at frequency $f$. 
Although actual seismic motion involves a broad spectrum of frequencies, we use $a$ as a proxy for net upward acceleration, assuming that overlapping modes constructively enhance vertical motion, as illustrated in Fig. 4b of \citet{Richardson2020}.

\subsection{Landslide toy model} \label{subsec:landslide}

We developed a toy model to estimate the downslope motion of a boulder on each facet of the simplified shape model over 10 minutes after the impact. 
We considered three forces: gravity, seismic acceleration, and friction.

Gravity was modeled using the Mascon method by dividing the interior of the asteroid into tetrahedra based on 3D Delaunay triangulation, implemented in the \texttt{scipy.spatial.Delaunay} module of the SciPy library \citep{2020SciPy-NMeth}, with a 10-m internal grid spacing. 
The total number of tetrahedra ($N_\mathrm{V}$) was 128\,672.

Previous studies reported a difference in internal density between the head and body regions of Itokawa \citep{Lowry2014,Kanamaru2019}. We adopted a density of $2\,450~\mathrm{kg\,m^{-3}}$ for the head ($X > +150~m$) and $1\,930~\mathrm{kg\,m^{-3}}$  for the body ($X < +150~m$) from \citet{Kanamaru2019}. We then calculated a composite force vector ($\mathbf{F}$) of gravity and centrifugal force acting on the centroid of each facet using the following equation:

\begin{equation}
    \mathbf{F} = G \sum_{\mathrm{i}=1}^{N_\mathrm{V}} 
    \frac{\rho_\mathrm{i} V_\mathrm{i}}{|\mathbf{r}_\mathrm{i} - \mathbf{r}|^3} 
    \left(\mathbf{r}_\mathrm{i} - \mathbf{r}\right) - \omega^2 \mathbf{r}_\perp,
    \label{eq:force}
\end{equation}

\noindent
where $G$ is the gravitational constant, $\rho_\mathrm{i}$, $V_\mathrm{i}$, and $\mathbf{r}_\mathrm{i}$ are the mass density, volume, and position vector of each volume element; $\mathbf{r}$ and $\mathbf{r}_\perp$ are the centroid’s position vector and perpendicular component from the spin axis; and $\omega$ is the spin rate. We used Itokawa’s rotational period of 12.1324 hours \citep{Fujiwara2006}. We defined the surface slope ($\theta$) as the angle between the facet normal and the gravity vector.

Seismic acceleration from Sect. \ref{subsec:propagation} was assumed to act along the normal direction of each facet. Since acceleration was computed for 10 seconds, we extrapolated the last 2 seconds over 10 minutes, assuming exponential decay. Lastly, we used friction coefficients ($\mu$) from \citet{DellaGiustina2024}, with 0.6 for both static and kinetic friction constants.

The motion of a surface boulder on each facet was integrated using the Euler method with a timestep of 0.025 seconds, based on the assumption that a mass undergoing harmonic motion experiences peak acceleration for approximately one-quarter of its oscillation period (the inverse of $f$ = 10 Hz). At each timestep, we first determined whether the boulder was located on or above the surface. If the boulder was above the surface, it was considered to be in ballistic motion and was influenced only by gravity. We assumed a completely inelastic collision, setting the boulder's velocity to zero upon impact with the slope surface. This assumption may underestimate subsequent motion after landing, such as rolling or bouncing. The implications of this assumption are discussed in Sect. \ref{subsec:uncertainty}.

If the boulder was on the slope, we compared the seismic acceleration $a$ to the normal component of gravitational acceleration, $g \cos \theta$. 
If $a > g \cos \theta$, the boulder was assumed to be launched and transitioned into ballistic motion. 
If $a < g \cos \theta$, we evaluated whether downslope sliding occurred by comparing the tangential gravitational force, $g \sin \theta$, to the frictional resistance, $\mu~(g \cos \theta - a)$. 
If the tangential force exceeded the frictional resistance, the boulder was considered to slide; otherwise, it remained stationary. We recorded the displacement of the boulder on each facet as a proxy for the extent of surface disturbance caused by seismic shaking.

\section{Results} \label{sec:result}

In this section, we present the numerical results derived from our two-step simulation framework. 
Section~\ref{result:acc} reports the results of the surface acceleration caused by seismic shaking, based on the diffusion model introduced in Sect.~\ref{subsec:propagation}. 
Section~\ref{result:slide} then evaluates the resulting surface displacement using the landslide toy model described in Sect.~\ref{subsec:landslide}.

\subsection{Surface acceleration by seismic wave}
\label{result:acc}

\begin{figure*}
\centering
 \includegraphics[width=\hsize]{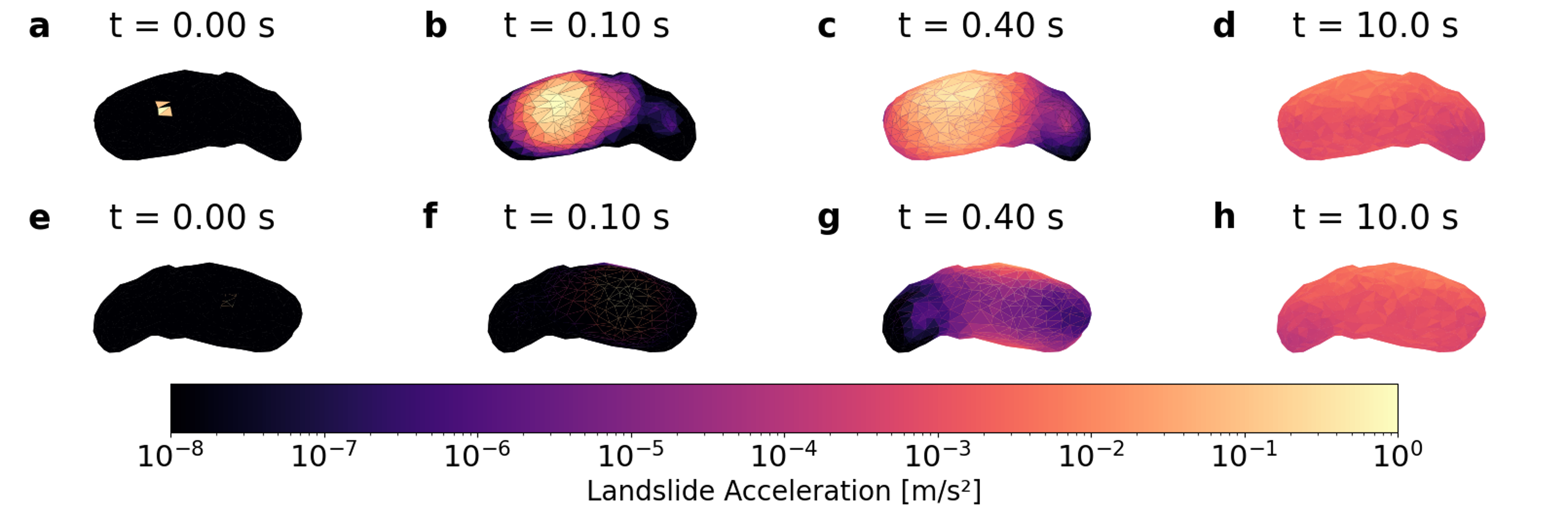}
\caption{Surface acceleration on the Kamoi (western) side (a-d) and the opposite (eastern) side (e-h) caused by seismic wave propagation at (a, e) 0 s (initial condition), (b, f) 0.1 s, (c, g) 0.4 s, (d, h) 10 s after the impact.}
\label{fig:propagation1}
\end{figure*}

%\begin{figure*}
%    \centering
%    \includegraphics[width=\hsize]{IS_propagation2_large.png}
%\caption{Surface acceleration on the opposite side, evaluated at the same time points as in Fig.~\ref{fig:propagation1}.}
%\label{fig:propagation2}
%\end{figure*}

All of our simulations show that surface acceleration varies over time as the seismic wave propagates through the asteroid. Figure \ref{fig:propagation1} visualizes acceleration changes for the nominal condition (Simulation 1 in Table \ref{table:sims}, $\eta = 1.0 \times 10^{-7}$ and $K$ = $3\,000 ~\mathrm{m^{2}\,s^{-1}}$) on the Kamoi (western) side (Fig.\ref{fig:propagation1}a–d) and the opposite (eastern) side (Fig.\ref{fig:propagation1}e–h). We found that the seismic energy causes high accelerations greater than $1~\mathrm{m\,s^{-2}}$ near the impact site within the first 0.10 seconds (Fig. \ref{fig:propagation1}a, b). 

At around 0.40 seconds after the impact, the seismic wave reaches the head part and the opposite (Eastern) side (Fig. \ref{fig:propagation1}c, g). For reference, $10^{-5}~\mathrm{m\,s^{-2}}$ is a typical surface gravitational acceleration on the Itokawa surface. Even after 10.0 seconds (see Fig. \ref{fig:propagation1}d, h), the surface acceleration remains above this typical surface gravitational acceleration ($a=10^{-5}~\mathrm{m\,s^{-2}}$) due to residual seismic energy, despite overall attenuation. 

\begin{figure*}
    \centering
    \includegraphics[width=\hsize]{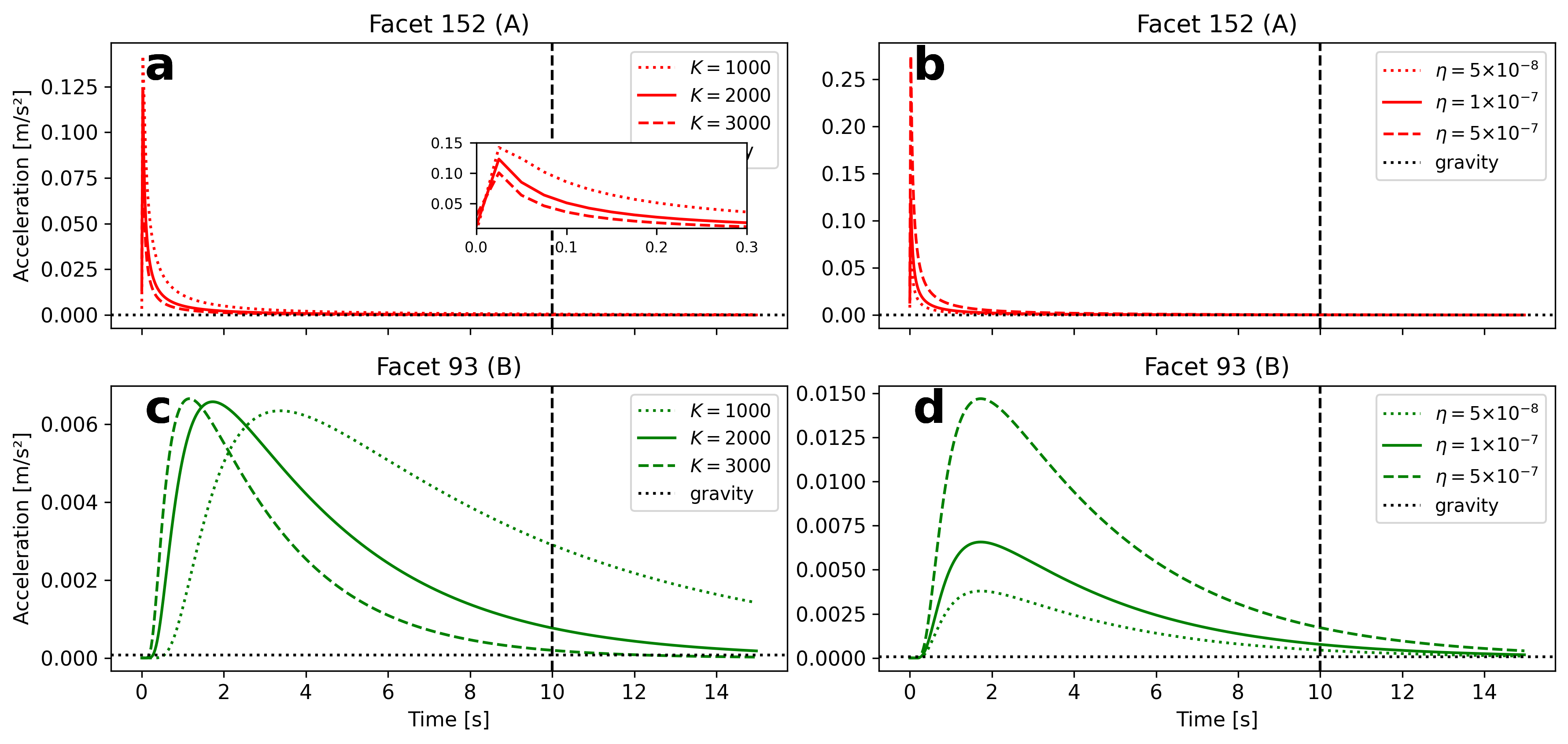}
\caption{
Time-varying acceleration experienced by each facet.  
(a) Acceleration at facet number 152 for different diffusivity values:  
$K = 1\,000~\mathrm{m^{2}\,s^{-1}}$ (Simulation 2, dotted),  
$K = 2\,000~\mathrm{m^{2}\,s^{-1}}$ (Simulation 1, solid), and  
$K = 3\,000~\mathrm{m^{2}\,s^{-1}}$ (Simulation 3, dashed).  
(b) Acceleration at facet number 152 for different seismic efficiency values:  
$\eta = 5.0 \times 10^{-8}$ (Simulation 5, dotted),  
$\eta = 1.0 \times 10^{-7}$ (Simulation 1, solid), and  
$\eta = 5.0 \times 10^{-7}$ (Simulation 8, dashed).  
(c, d) Same as (a, b), respectively, but for facet number 93.  
The horizontal dotted line represents the gravitational acceleration at each facet.  
The vertical dashed line marks $t = 10$ seconds, when the simulation ends and exponential decay extrapolation begins.
}
    \label{fig:acceleration}
\end{figure*}

Figure~\ref{fig:acceleration} shows the dependency of surface acceleration on $K$ and $\eta$ for facets A and B (defined in Fig.~\ref{fig:simplify}) with facet A located near and facet B farther from the Kamoi crater, respectively. Both facets experience accelerations one order of magnitude higher than the local gravitational acceleration (indicated by the horizontal dotted line). Since facet A is close to the impact site, its peak acceleration is higher than that of facet B and appears within 0.1 seconds after the impact. 
In contrast, facet B shows its peak after around 1 second under the highest diffusivity condition ($K = 3\,000~\mathrm{m^{2}\,s^{-1}}$, dashed lines). In addition, facet A exhibits both a higher peak acceleration and a longer duration above the gravitational threshold when $K$ is low (dotted line) or $\eta$ is high (red dashed line). 
Conversely, facet B shows stronger acceleration under higher $K$ (green dashed line) and $\eta$ (green dotted line), although a larger $K$ leads to faster attenuation, reducing the time during which the acceleration exceeds gravity.

\begin{figure*}
\centering
 \includegraphics[width=\hsize]{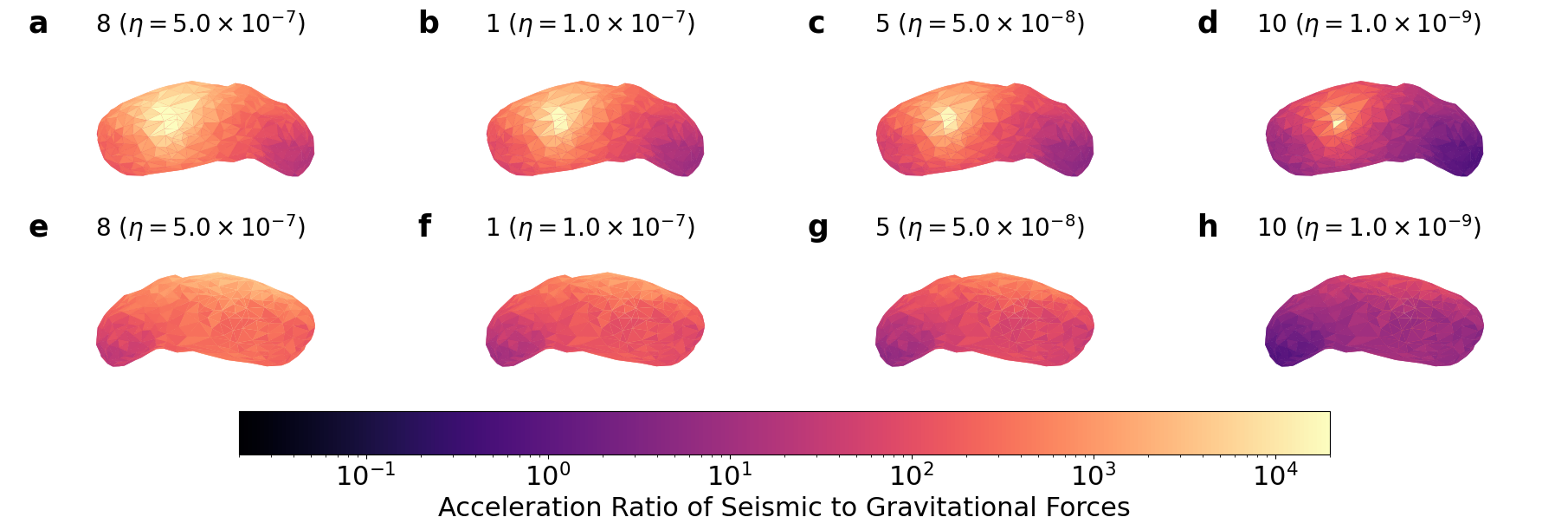}
\caption{
Ratio of maximum surface acceleration to local gravitational acceleration on the Kamoi side (Western; panels a–d) and the opposite side (Eastern; panels e–h) for different $\eta$ values. All simulations assume a constant $K$ value of 2,000. Panels are arranged such that $\eta$ decreases from left to right. Each column corresponds to an $\eta$ value of $5\times10^{-7}$, $1\times10^{-7}$, $5\times10^{-8}$, and $1\times10^{-9}$, respectively.
}
\label{fig:amax1}
\end{figure*}

%\begin{figure*}
%    \centering
%    \includegraphics[width=\hsize]{IS_amax_ratio2_large.png}
%    \caption{Same as Fig. \ref{fig:amax1}, but for the opposite side.} 
%    \label{fig:amax2}
%\end{figure*}

\begin{figure}
    \centering
    \includegraphics[width=\hsize]{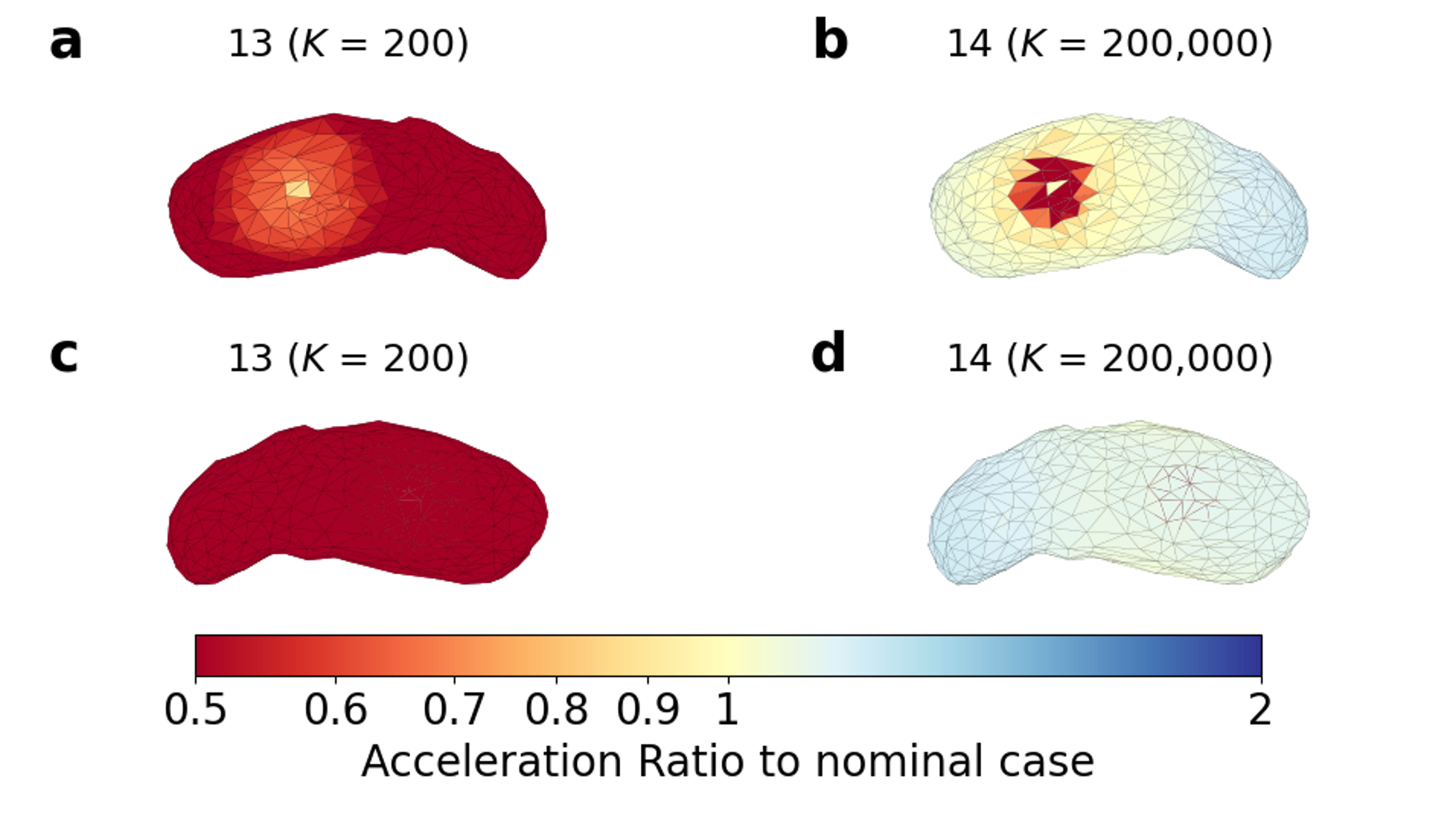}
    \caption{Ratio of maximum surface acceleration from simulations with extremely low ($K = 200~\mathrm{m^{2}\,s^{-1}}$, panels a, c) and high ($K = 200\,000~\mathrm{m^{2}\,s^{-1}}$, panels b, d) diffusivity values with respect to the nominal case. Panels (a) and (b) show the Kamoi side, while panels (c) and (d) show the opposite side.}  
    \label{fig:amax_extreme}
\end{figure}

Figure~\ref{fig:amax1} shows the ratio between the maximum acceleration during the 10-second simulation and the local surface gravity for the Kamoi (Western) side and the opposite (Eastern) side. 
For both sides, the ratio exceeds 0.2 across all facets, a value previously suggested as a threshold for boulder destabilization \citep{Miyamoto2014}. 
Panels~a and e in Figure~\ref{fig:amax1}, which correspond to larger $\eta$ values, show higher seismic accelerations compared to panels d and h with smaller $\eta$ values. 
In Simulation~10 ($\eta = 1.0 \times 10^{-9}$), the minimum acceleration ratio reaches approximately 0.2 at the facet farthest from the Kamoi crater (Fig.~\ref{fig:amax1}d, h).

Compared to $\eta$, the effect of changing $K$ is less prominent. To clarify the influence of diffusivity, we compared maximum accelerations from simulations with extreme $K$ values to the nominal case in Fig. \ref{fig:amax_extreme}. 
The lower $K$ cases (panels a, c) yield smaller peak accelerations than the nominal case, while the higher $K$ cases (panels b, d) produce greater accelerations across most surface regions, except near the Kamoi crater.

\subsection{Landslide simulation}
\label{result:slide}

\begin{figure*}
    \centering
    \includegraphics[width=\hsize]{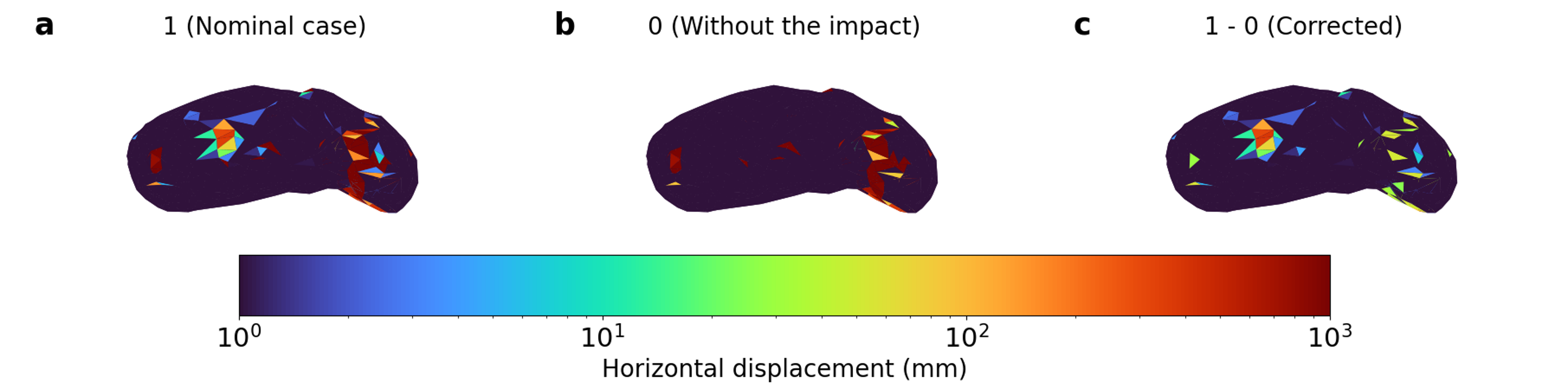}
    \caption{Correction process applied to the landslide simulation results: (a) result from the simulation under the nominal case,  (b) result from the simulation without seismic acceleration, and (c) difference between (a) and (b), used as the corrected displacement.
    }
    \label{fig:x_correction}
\end{figure*}

Figure \ref{fig:x_correction}a shows horizontal displacements estimated using our landslide toy model. We observed that some facets exhibited large displacements even in the absence of seismic acceleration from the Kamoi impact (Fig.~\ref{fig:x_correction}b). This is attributed to steep local slopes exceeding the angle of repose, which is 31.0 \degr for $\mu = 0.6$. To isolate the effect of seismic shaking, we corrected for this by subtracting the results of a control simulation without seismic input from those of the nominal simulation (Fig. \ref{fig:x_correction}c). Even after this correction, facets with steep slopes still show horizontal displacements of approximately 10 mm. 

\begin{figure*}
    \centering
    \includegraphics[width=\hsize]{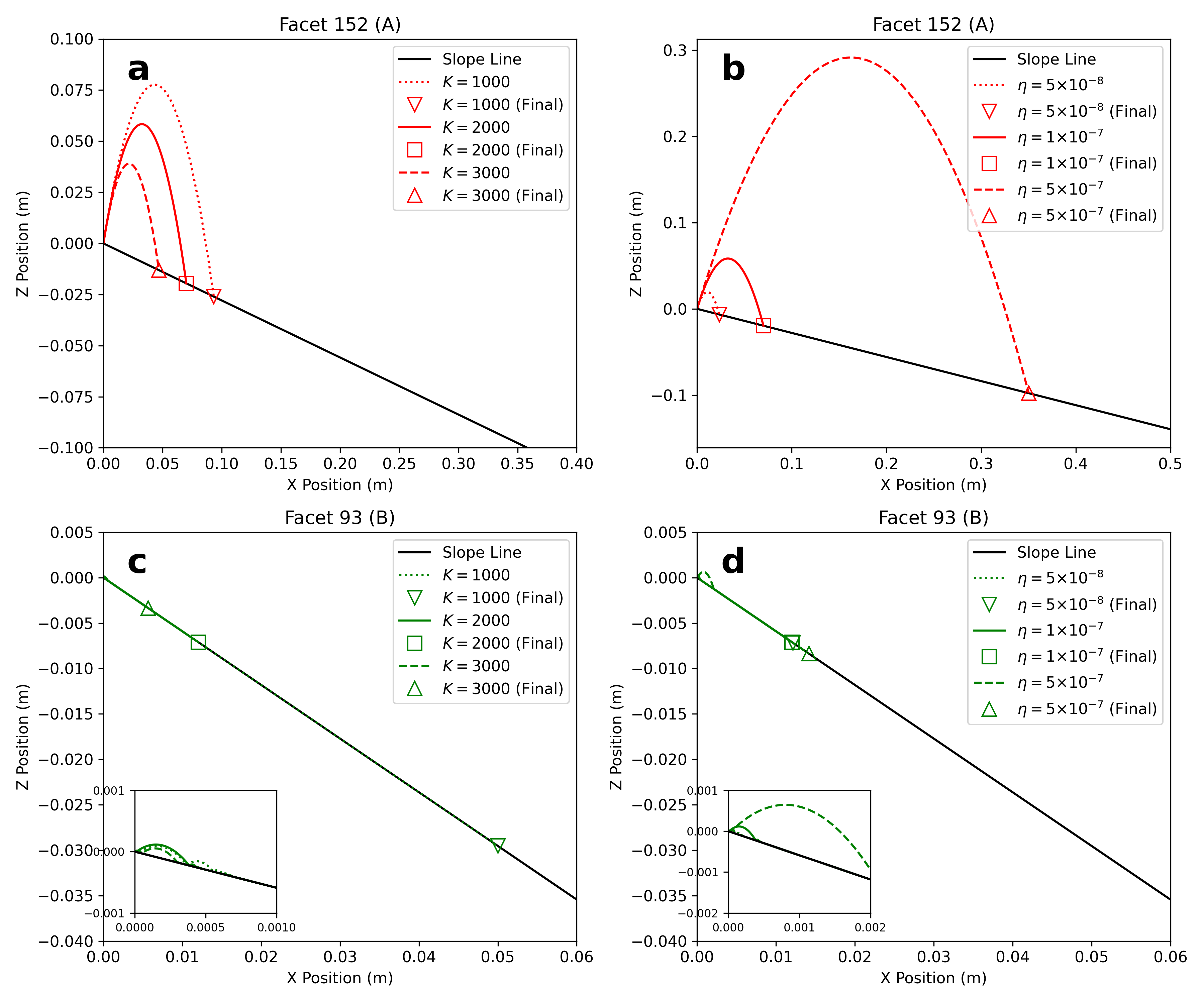}
\caption{Motion of boulders on each facet.  (a) Boulder motion on facet number 152 for different diffusivity values: $K = 1\,000~\mathrm{m^{2}\,s^{-1}}$ (Simulation 2, dotted line with downward triangle), $K = 2\,000~\mathrm{m^{2}\,s^{-1}}$ (Simulation 1, solid line with square), and $K = 3\,000~\mathrm{m^{2}\,s^{-1}}$ (Simulation 3, dashed line with upward triangle). (b) Boulder motion on facet number 152 for different seismic efficiency values: $\eta = 5.0 \times 10^{-8}$ (Simulation 5, dotted line with downward triangle),  $\eta = 1.0 \times 10^{-7}$ (Simulation 1, solid line with square), and $\eta = 5.0 \times 10^{-7}$ (Simulation 8, dashed line with upward triangle). (c) and (d) show the same comparisons as (a) and (b), respectively, but for facet number 93. Markers indicate the final positions of the boulders 10 minutes after the impact. Black inclined lines represent the local surface slope. }
    \label{fig:x_facet}    
\end{figure*}

The motion of boulders on facets 152 (A) and 93 (B) is shown in Fig. \ref{fig:x_facet}. A boulder on facet 152 is lifted from the surface due to strong initial acceleration and remains at its landing position (Fig. \ref{fig:x_facet}a, b), as the seismic acceleration is significantly attenuated by diffusion after 2 seconds (Fig. \ref{fig:acceleration}a, b). Larger horizontal displacements and higher maximum heights occur under low $K$ and high $\eta$ conditions. On the other hand, in Fig.~\ref{fig:x_facet}c and d, boulders undergo both launching and sliding due to seismic shaking. The case with $K = 1\,000~\mathrm{m^{2}\,s^{-1}}$ (green dotted line) shows greater displacement due to the long-lasting acceleration above gravity (dotted line from Fig. \ref{fig:acceleration}c). However, no clear relationship between $\eta$ and displacement is found in Fig.~\ref{fig:x_facet}d.

\begin{figure}
\centering
\includegraphics[width=\hsize]{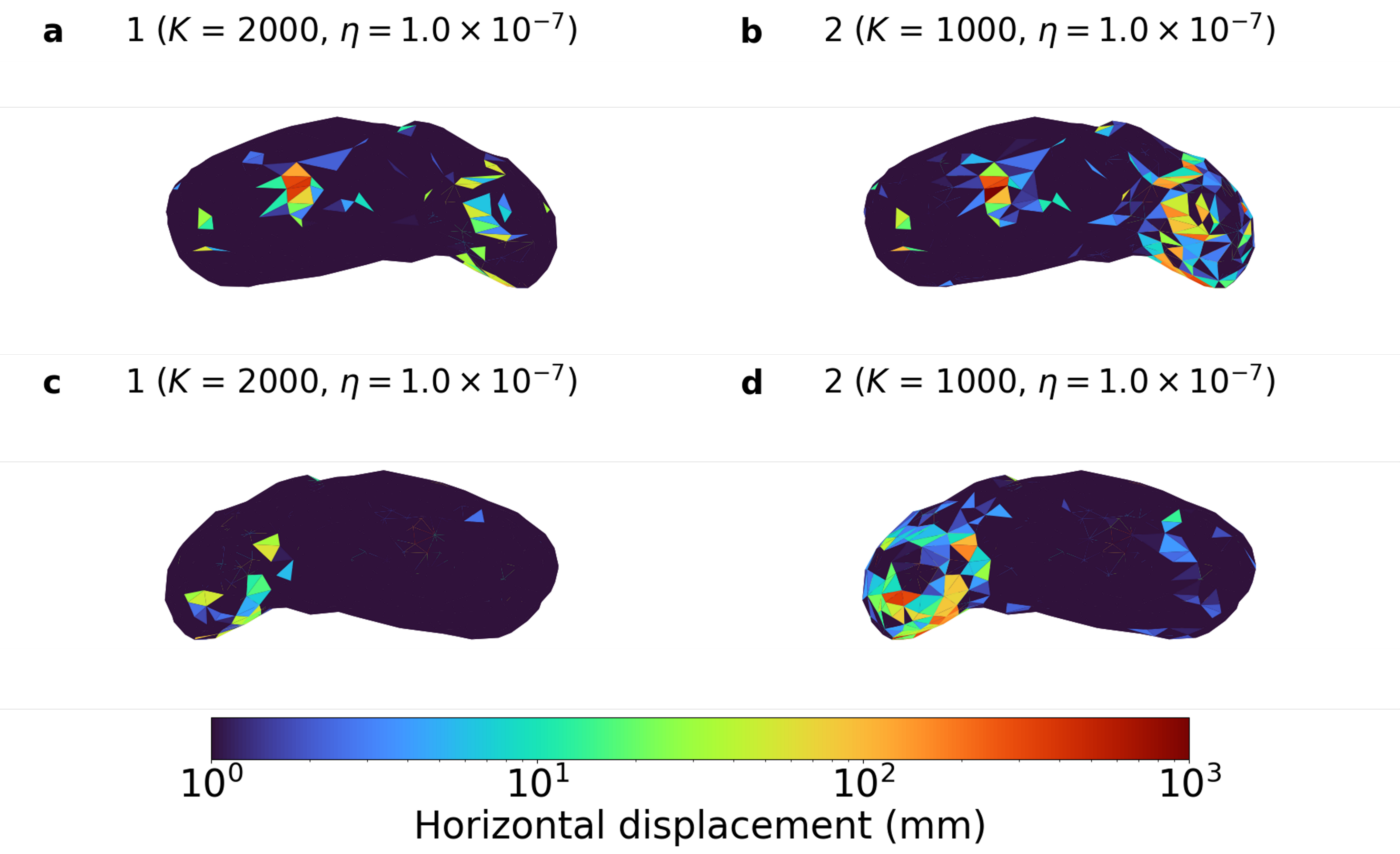}
\caption{Horizontal displacement of boulders on the Western (Kamoi) side (panels a, b) and the Eastern (opposite) side (panels c, d). Panels a and c show results from Simulation 1 ($\eta = 1.0 \times 10^{-7}$, $K = 2,000~\mathrm{m^{2}\,s^{-1}}$), while panels b and d are from Simulation 2 ($\eta = 1.0 \times 10^{-7}$, $K = 1,000~\mathrm{m^{2}\,s^{-1}}$), which best reproduces the observed space weathering distribution on Itokawa.} 

%Additional simulation results are provided in the Supplementary Material. 
%\textcolor{red}{how can we add supplementary figs?}}\textcolor{blue}{[Question: is this a question from Sunho to me (Ishiguro)? To include supplementary figures in the manuscript, you can use the \texttt{\textbackslash begin\{appendix\}} and \texttt{\textbackslash end\{appendix\}} environment. Figures placed within this environment will appear as part of the Appendix section.}

\label{fig:x1}
\end{figure}

\begin{figure*}
    \centering
    \includegraphics[width=\hsize]{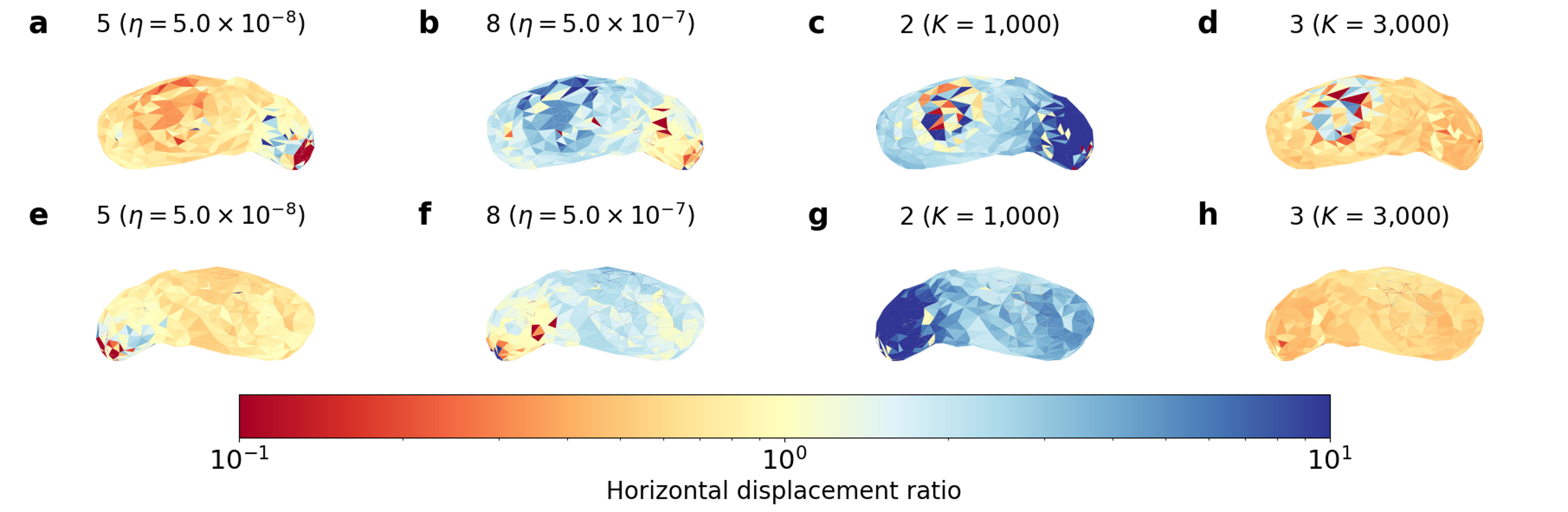}
    \caption{Horizontal displacement ratio relative to the nominal case (Simulation 1). Panels a–d show the Kamoi (western) side, and panels~e–h show the opposite (eastern) side. The first two columns present simulation results with $\eta$ values of $5.0 \times 10^{-8}$ (a, e) and $5.0 \times 10^{-7}$ (b, f), respectively. The last two columns show simulation results with $K$ values of $1\,000~\mathrm{m^{2}\,s^{-1}}$ (c, g) and $3\,000~\mathrm{m^{2}\,s^{-1}}$ (d, h).}
    \label{fig:x_ratio}
\end{figure*}

\begin{figure}
    \centering
    \includegraphics[width=\hsize]{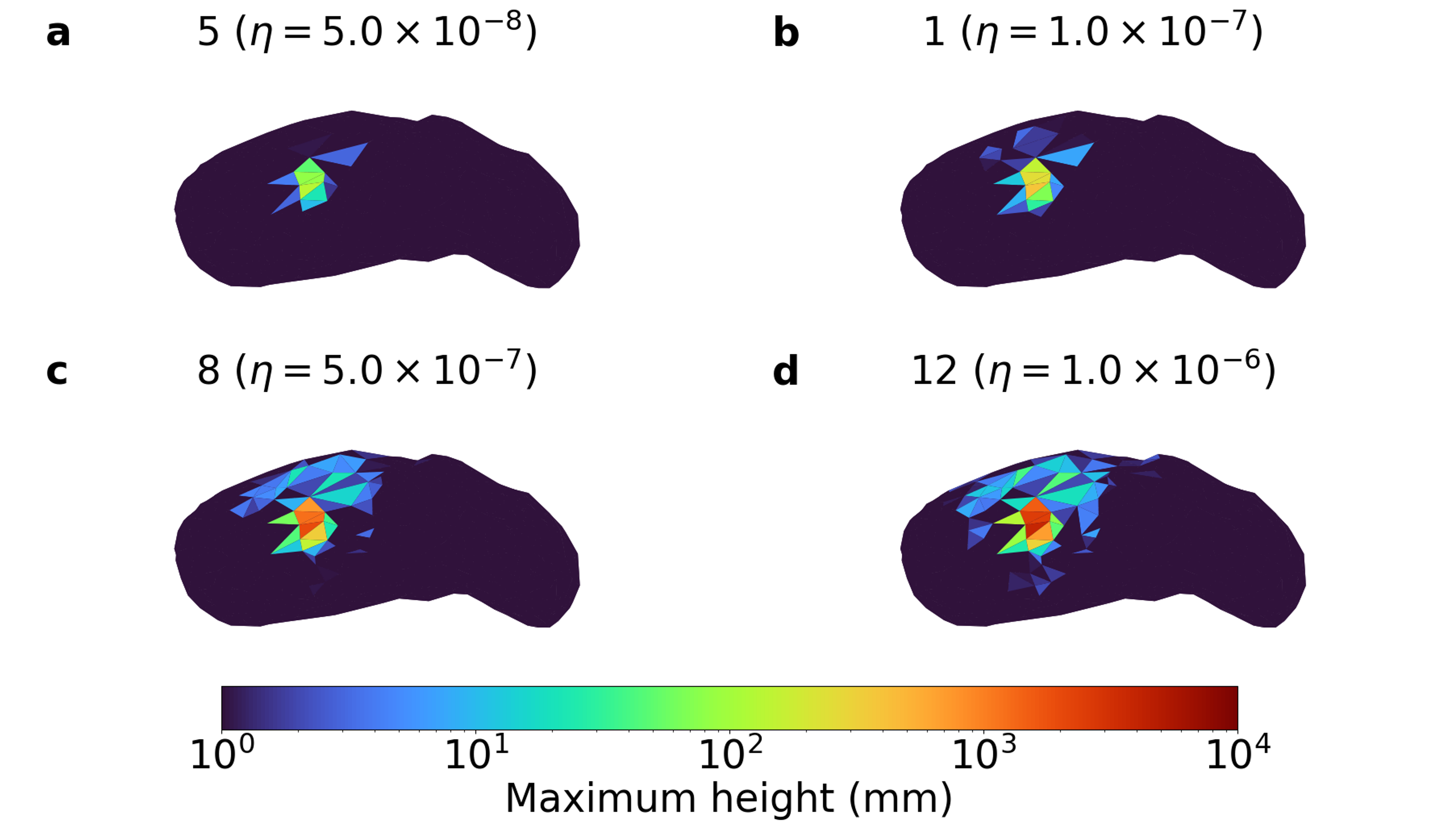}
    \caption{Maximum vertical displacement of boulders during the simulation on the Kamoi (Western) side. Panels (a–d) correspond to $\eta$ values of $5 \times 10^{-8}$, $1 \times 10^{-7}$, $5 \times 10^{-7}$, and $1 \times 10^{-6}$, respectively. In panels~(a–c), the maximum displacement ranges from 1~cm to 1~m, whereas in panel~(d), it exceeds 1~m.}
    \label{fig:z1}
\end{figure}

Representative simulation results using $K = 1\,000$ and $2\,000~\mathrm{m^{2}\,s^{-1}}$ and $\eta = 1 \times 10^{-7}$ from \citet{Richardson2020} are shown in Fig.~\ref{fig:x1}. For the Kamoi side (Fig.~\ref{fig:x1}a, b), regions near the Kamoi crater, steep areas around the neck connecting the body (left) and head (right) lobes of Itokawa, and some locally steep facets exhibit horizontal displacements greater than 1~mm. Meanwhile, the opposite side (Fig.~\ref{fig:x1}c, d) shows lower overall displacements. Nevertheless, similar to the Kamoi side, some steep-sloped facets (particularly those near the neck) exhibit displacements on the order of 1~cm. 
In the case of $K = 1\,000~\mathrm{m^{2}\,s^{-1}}$ (Figs.~\ref{fig:x1}b, d), the head lobe experiences significant horizontal motion on both sides. This trend is clearly visible in Figs.~\ref{fig:x_ratio}c and g, where the head region shows displacements more than ten times greater than those in the surrounding areas. In contrast, the high-$K$ case (Figs.~\ref{fig:x_ratio}d, h) exhibits generally low displacements, except in regions near Kamoi. The high-$\eta$ case (Figs.~\ref{fig:x_ratio}a, e) shows elevated displacement values across the surface compared to the low-$\eta$ case (Figs.~\ref{fig:x_ratio}b, f), except at some facets on the head lobe.
In addition, Fig.~\ref{fig:z1} shows the maximum vertical height reached during the landslide on the Kamoi side. Launch heights greater than 1 mm appear only near the Kamoi crater and increase with higher $\eta$.

\section{Discussion}
In this section, we first examine the modeling assumptions and evaluate how they may influence the results of our seismic and landslide simulations (Sect.~\ref{subsec:uncertainty}). We then compare the simulation outcomes with the observed distribution of space weathering on Itokawa and discuss the implications of the inferred seismic diffusivity ($K$) and efficiency ($\eta$) for understanding resurfacing processes on small asteroids.

\subsection{Assumptions and Uncertainties in Seismic and Landslide Modeling} \label{subsec:uncertainty}
To make the modeling manageable, we adopted several simplifying assumptions in simulating both seismic wave propagation and landslides. In the following subsections, we evaluate how these assumptions might influence our results.

\subsubsection{Seismic energy propagation} \label{subsubsec:uncertainty_energy}

We applied the Neumann boundary condition, assuming that the vacuum outside the asteroid exerts no external stress, when we solved Eq.~\ref{eq:diffusion}. 
This assumption neglects energy loss through launching boulders or regolith particles, a process known as the “cocoa effect” \citep{Tancredi2023}. These losses could slightly reduce surface accelerations. However, since this process also contributes to surface renewal, omitting it may not undermine our conclusions and might also contribute to resurfacing.

We modeled the asteroid surface as a simple harmonic oscillator with a single frequency of 10~Hz. This value was chosen based on the dominant spectral peak around 10–-20~Hz reported by \citet{Richardson2020}. This frequency is also consistent with impact-generated seismic waves observed on the Moon \citep{Latham1973, Duennebier1974}. Although 10~Hz likely represents the dominant frequency generated by the Kamoi impact, real seismic waves typically contain a broad frequency spectrum.

To test the robustness of this assumption, we performed additional simulations at 1~Hz and 100~Hz. Higher-frequency simulations result in faster energy attenuation due to the larger attenuation term in Eq.~\ref{eq:diffusion}, leading to surface accelerations lower by a factor of about two for the 100~Hz case. Despite this, near-crater peak accelerations remained similar across all frequencies. Moreover, the overall horizontal displacement varied by less than an order of magnitude for all cases, and the locations of significantly altered terrain remained unchanged. These results suggest that the model’s predictions of surface modification are relatively insensitive to the assumed frequency, reinforcing the robustness of our conclusions.

Additionally, \citet{Quillen2019} estimated the corner frequency ($f_\mathrm{c}$)—the cutoff above which the seismic source spectrum rapidly weakens—using the relation:

\begin{equation}
f_\mathrm{c} = \frac{v_\mathrm{p}}{D_\mathrm{crater}},
\end{equation}

\noindent where the P-wave velocity $v_\mathrm{p}$ is $100~\mathrm{m\,s^{-1}}$ and the crater diameter $D_\mathrm{crater}$ is 8~m \citep{Yasui2015, 2009Icar..200..486H}. This yields a corner frequency of 12.5~Hz, suggesting that significantly higher frequencies contribute little to the seismic response, further supporting our 10~Hz assumption.

Furthermore, we fixed a quality factor ($Q$) in Eq.~\ref{eq:diffusion} to 1\,500. We adopted this value because $Q$ values estimated by \citet{Richardson2020} for asteroids, Eros, Steins, and Itokawa, range from 1500 to 2000, although their sizes are significantly different from each other. To evaluate the sensitivity of $Q$ on the final result, we also ran simulations with $Q = 500$ and $4\,000$, and confirmed that the distribution and amount of horizontal displacements varied by less than an order of magnitude. In addition, since the attenuation term in Eq.\ref{eq:diffusion} ($2\pi f \epsilon / Q$) includes both frequency and $Q$, the effects of frequency and $Q$ can partially cancel out. Given that our model already showed robust results over a wide frequency range (1-–100 Hz), the variation in $Q$ is unlikely to significantly affect the conclusions.

Moreover, we modeled the propagation of seismic waves as a diffusion process, rather than as coherent wave propagation, following \citet{Richardson2005}. According to \citet{Nakamura1977}, seismic waves may be treated as particles traveling through a scattering medium if the condition

\begin{equation}
    D \geq k \lambda
    \label{eq:diffusion_condition}
\end{equation}

\noindent is satisfied, where $D$ is the characteristic size of scatterers, $k$ is a constant (typically $k \geq 1$), and $\lambda$ is the wavelength.

In our nominal case ($K = 2\,000~\mathrm{m^{2}\,s^{-1}}$), assuming a wave velocity of $100~\mathrm{m\,s^{-1}}$ derived from impact experiments on highly porous materials \citep{Yasui2015}, the mean free path $l$ of the wave is determined to be 60~m from Eq.~\ref{eq:diffusivity}, which may be considered as $D$. On the other hand, $\lambda$ is derived to be 10~m from the assumed velocity and frequency. This yields $k = 6$, which satisfies the condition for the diffusion approximation. Therefore, it is safe to say that this is a self-consistent simulation to simplify seismic wave propagation using the diffusion equation given by Eq.~\ref{eq:diffusion}.

Although our simulation results indicate that Itokawa has lower seismic diffusivity values ($K$) than other asteroids \citep{Richardson2020}, reflecting its highly scattering interior nature, our model does not explicitly account for wave reflections and scattering at interfaces between interior blocks within Itokawa. This simplification may lead to either overestimation or underestimation of seismic energy transport. A more advanced approach, such as the smooth sphere discrete element method, could better capture these complexities and could be explored in future studies.

\subsubsection{Landslide toy model} \label{subsubsec:uncertainty_landslide}
We used a low-resolution shape model to eliminate effects of boulder-scale topography present in the original shape model \citep{Gaskell2020}. This simplification may overlook small-scale landslides on steep local slopes, particularly those smaller than a few tens of meters in scale. As a result, our model cannot reproduce localized landslides, such as those occurring at crater rims or along ridge flanks. We acknowledge this limitation and revisit it in Sect.~\ref{subsec:itokawa_sw}.

Additionally, we assumed that acceleration acts only along the surface normal, which may neglect lateral shaking effects. The simple harmonic oscillator assumption may also become less accurate when the mass of a boulder is comparable to the local substrate volume. In such cases, the boulder and underlying surface behave as a coupled mass system rather than a driven oscillator, reducing relative acceleration and displacement. This effect is expected to be minor, however, since only six boulders larger than 20~m have been identified on Itokawa \citep{Michikami2008}.

We adopted a constant coefficient of friction, $\mu = 0.6$, for both static and kinetic friction, following \citet{DellaGiustina2024}. To assess the sensitivity of our results, we also tested a case with a higher friction coefficient of $\mu = 1.0$. In this case, the angle of repose increased from 31 \degr to 45 \degr, reducing the number of facets that experienced sliding. Nevertheless, the overall distribution of horizontal displacements remained similar to the nominal case, suggesting that our results are robust with respect to variations in the friction coefficient.

\begin{figure*}
\centering
\includegraphics[width=\hsize]{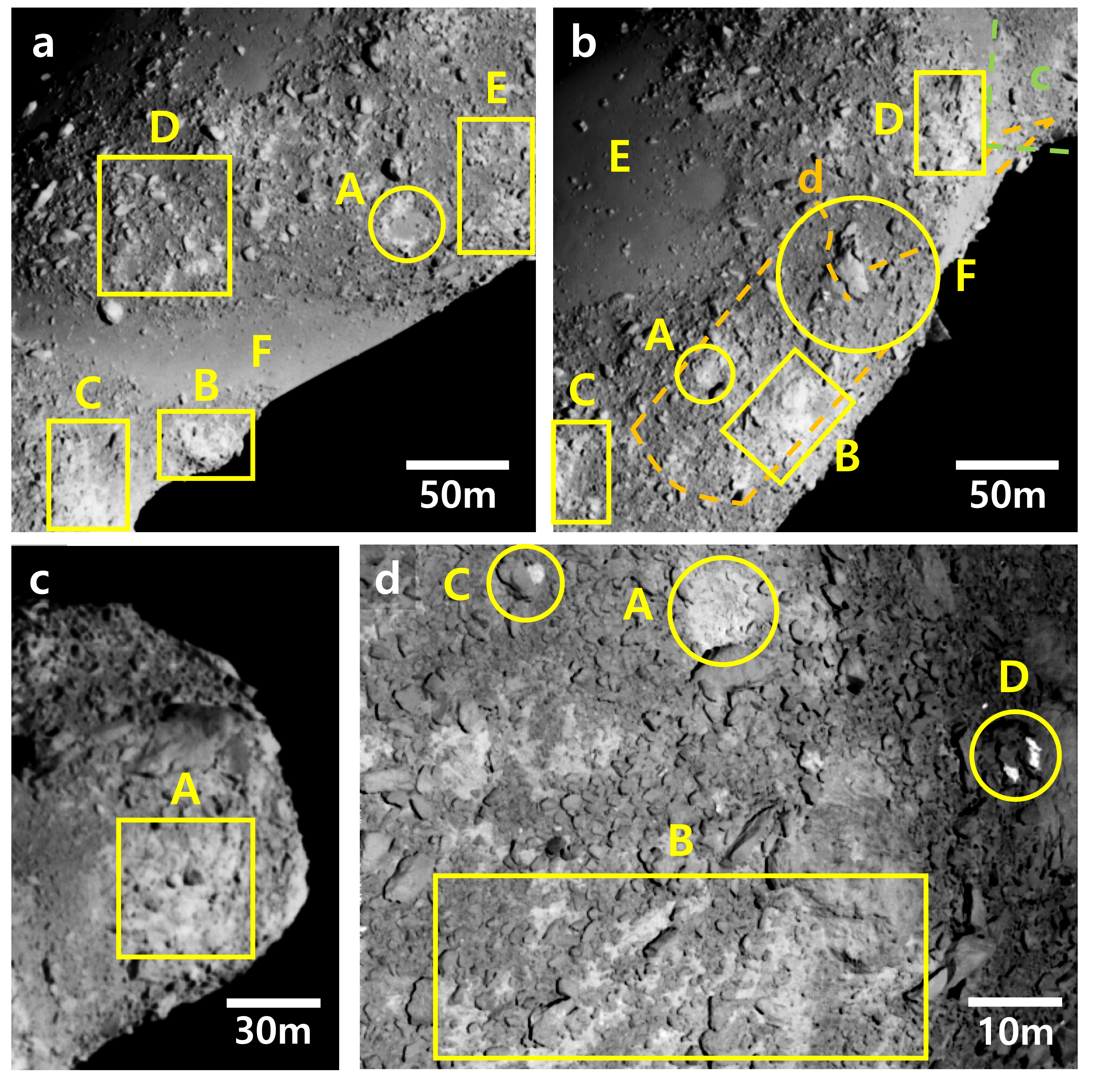}
\caption{
High-resolution images of Itokawa taken by Hayabusa/AMICA. 
(a) Eastern (opposite) side of Itokawa (image ID: ST\_2487335302). This image includes the Komaba crater with a bright rim (A), fresh surfaces on steep slopes along the Yatsugatake ridge (B), Shirakami (C), and other fresh patches near large boulders or craters (D, E). The large smooth terrain, Muses Sea, is also shown (F). 
(b) Western (Kamoi) side (image ID: ST\_2486649845). Locations of panels (c) and (d) are marked with orange and green dashed lines, respectively. Notable features include the Kamoi crater (A), fresh surfaces near the Kamoi crater (B, C) and in the neck region (D), the mature smooth terrain Sagamihara Planitia (E), and another smooth area near the Tsukuba boulder (F). 
(c) Head region (image ID: ST\_2483892216), showing relatively fresh terrain near a black boulder (A). 
(d) Close-up near the Kamoi crater (image ID: ST\_2566571276), showing the crater (A), fresh basins around weathered boulders (B), and boulders with very fresh surfaces likely fractured by impact or thermal fatigue (C, D).
}
\label{fig:detail}
\end{figure*}

Finally, we ignored the rolling and bouncing of boulders by assuming perfectly inelastic impacts upon landing. This may lead to an underestimation of the final horizontal travel distance. However, dark boulder surfaces even at apparently fresh terrains (Fig.~6b of \citealt{Ishiguro2007}) imply long surface exposure ages and launch heights of less than 1~m. This suggests that secondary motion is limited on Itokawa.

Despite the simplifications described in this subsection (i.e., the use of a low-resolution shape model, the neglect of lateral shaking and secondary motions, and the assumption of purely inelastic boulder impacts), our simulations successfully reproduce the major patterns of global seismic shaking and landslides observed on both the western and eastern hemispheres. These assumptions generally tend to underestimate the extent of landslides. Therefore, our argument that the Kamoi impact triggered global surface renewal is likely robust. The reproduction of key observational features under such limiting conditions supports the validity of our approach and strengthens our overall conclusions.

\subsection{Comparison to space weathering distribution of Itokawa} \label{subsec:itokawa_sw}

Based on observed spatial variations in space weathering, we summarize the key observational features as follows:

\begin{enumerate}
    \item The Kamoi (western) side exhibits lower space weathering degrees than the opposite (eastern) side \citep{Ishiguro2007}.
    \item Fresh, high-albedo patches are observed around large boulders near the Kamoi crater and in the head region, likely indicating recently exposed surface material (Fig.~\ref{fig:detail}b, d).
    \item Relatively fresh terrains are also appear in localized regions on the opposite hemisphere, particularly around the southern neck and the rim of the Komaba crater (Fig.~\ref{fig:detail}a: A, C; b: D).
    \item The smooth terrain Muses Sea tends to show intermediate space weathering degrees (Fig.~\ref{fig:detail}a: F).
\end{enumerate}

In the following subsections, we compare these observed features with the results of our seismic shaking simulations and discuss possible interpretations.

\subsubsection{Space weathering dichotomy between Kamoi and opposite sides}

In this subsection, we assess whether our simulation results reproduce the first key observational feature: (1) the western (Kamoi) side (Fig.~\ref{fig:detail}b) exhibits lower space weathering degrees than the opposite side (Fig.~\ref{fig:detail}a). Our simulation shows that maximum accelerations are systematically higher on the Kamoi side (e.g., Fig.~\ref{fig:propagation1}a-d) than on the opposite side (Fig.~\ref{fig:propagation1}e-h). As a result, surface materials on the Kamoi side experience larger horizontal displacements, increasing the likelihood that fresh subsurface materials are exposed over a broad region, not only near the impact site (Fig. \ref{fig:x1}a, b). This trend agrees with the observational evidence reported by \citet{Ishiguro2007}, who reported that the Kamoi side is less weathered overall.

Hayabusa/AMICA observations also reveal terrains with intermediate space weathering degrees. Our simulation indicate that these areas correspond to horizontal displacements of less than approximately 1~m. We propose that the intermediate space weathering signatures result from spatial and spectral mixing between fresh and matured surfaces, causing these terrains to appear only partially weathered in remote-sensing data.

This hemispherical dichotomy also appears inconsistent with the hypothesis that thermal fatigue is the dominant resurfacing mechanism on Itokawa. Although we identified features that may have originated from boulders recently broken by thermal fatigue (Fig.~\ref{fig:detail}d: C, D), such occurrences are rare. The spin axis of Itokawa is oriented at $(\lambda, \beta) = (128.5^\circ, 89.66^\circ)$ \citep{Demura2006}, nearly perpendicular to its orbital plane. This orientation results in similar diurnal heating and cooling cycles across both hemispheres, making it difficult to explain the observed dichotomy in space weathering degrees through thermal fatigue alone \citep{Delbo2014}.

% \subsubsection{Low space weathering degrees at rough terrains}
\subsubsection{Fresh surface exposures around large boulders}

Close-up images (Fig.~\ref{fig:detail}b: B,C; d: B) reveal bright, high-albedo patches appear immediately adjacent to large boulders in some boulder-rich regions, particularly near the Kamoi crater and the head part. These features are interpreted as exposures of the original substrate where the boulders were once seated, which became freshly exposed due to seismic shaking. The elevated reflectance of these patches suggests that the subsurface material was less weathered, and the motion of boulders helped uncover these fresh surfaces. Most of these bright patches on the western (Kamoi) side are reproduced in our simulations (Fig.~\ref{fig:x1}a, b). For example, around the Kamoi crater, including its upper-left (Fig. \ref{fig:detail}b: C) and lower (Fig. \ref{fig:detail}b: B) portions, leading to boulders being displaced and redeposited atop others, thereby exposing the previously buried substrate.

We estimate that a minimum launching height of approximately 1 cm, comparable to Itokawa’s characteristic grain size, is sufficient to expose fresh material \citep{Miyamoto2007}. However, we also infer that the maximum launching height does not exceed about 1 meter, as most boulders retain high space weathering degrees, suggesting that they were not significantly overturned. Based on this constraint, we narrow the plausible range of seismic efficiency $\eta$ to between $5.0 \times 10^{-8}$ and $5.0 \times 10^{-7}$, as shown in Fig.~\ref{fig:z1}. This range is consistent with the estimates of \citet{Richardson2020} for Itokawa. 

\subsubsection{Localized fresh terrains on the opposite hemisphere}

Even on the hemisphere opposite to the Kamoi side (eastern side), we observe small regions where fresh subsurface material appears to have been exposed, likely due to landslides. In particular, relatively bright terrains have been identified in high-slope areas such as the southern neck region (Fig.~\ref{fig:detail}b: D) and along the rim of the Komaba crater (Fig.~\ref{fig:detail}a: A).

Our simulation does not reproduce such localized landslides on steep slopes, mainly due to the use of a simplified shape model with reduced spatial resolution. Even with a higher-resolution model, it would be challenging to capture landslides in such small-scale, steep terrains. Nevertheless, if slopes around the crater rim are inclined at approximately 30 \degr, our model suggests that they could become unstable under the modeled seismic conditions. We therefore conjecture that the bright terrains observed near the Komaba crater rim and the southern neck may also have resulted from seismic shaking due to the Kamoi impact.

Simulations with lower $K$ values produce significant displacements in the head region, driven by prolonged accelerations that exceed the gravitational threshold (see Fig.~\ref{fig:acceleration}). These displacements correspond well with observed fresh terrains (Fig.~\ref{fig:detail}c: A), suggesting that $K$ likely falls within the range of 1\,000 to 2\,000~$\mathrm{m^2\,s^{-1}}$. This $K$ estimate is consistent with that of \citet{Richardson2020}.

In summary, our results support the hypothesis that the Kamoi impact induced global seismic shaking, triggering landslides even on the opposite hemisphere and explaining the distribution of fresh terrains across both sides of Itokawa.

\subsubsection{Moderate space weathering degrees in Muses Sea}
Smooth terrains on Itokawa, such as Muses Sea (Fig.~\ref{fig:detail}a: F), tend to exhibit moderate degrees of space weathering. Pre-landing images revealed that these areas are covered with fine-grained regolith composed of millimeter- to centimeter-sized particles \citep{Yano2006}. These fine particles are susceptible to granular convection, a process that stirs the regolith vertically and horizontally, and redistributes materials from different depths \citep{Miyamoto2007}. As a result, the continuous mixing of mature and relatively fresh particles may lead to an intermediate spectral signature, appearing as moderately weathered terrain in remote sensing data.

This interpretation is also consistent with the laboratory analyses of Itokawa samples, which were mostly collected from the Muses Sea region. These sampled particles exhibit space-weathered rims on the examined grains \citep{Matsumoto2015}, suggesting that regolith mixing through granular convection allowed even subsurface particles to be exposed and weathered over time. We thus infer that the observed moderate space weathering degrees are a result of long-term regolith turnover extending to depths of a few meters, leading to a spatially averaged weathering state across the smooth terrains \citep{Yano2006}.

In summary, our simulation results are broadly consistent with the observed space weathering patterns on Itokawa. All key features can be explained by seismic shaking and landslides triggered by the Kamoi impact. Only small-scale features remain unresolved due to the shape model resolution, but the overall spatial trends are well reproduced.

\subsection{Unique space weathering properties of Itokawa among S-type asteroids}
\label{subsec:uniqueness}

Itokawa is unique among S-type near-Earth asteroids visited by spacecraft, such as (433) Eros, (4179) Toutatis, and (65803) Didymos, as fresh surface exposures are widespread across its surface. In addition, other S-type asteroids in the main belt, (951) Gaspra and (243) Ida, exhibit globally weathered, matured surfaces. Here, we consider the origin of Itokawa’s unusual surface freshness.

Figure~\ref{fig:amax1} shows that the maximum surface acceleration exceeded the conventional threshold for boulder destabilization (0.2; \citealt{Miyamoto2014}) across the entire surface by at least an order of magnitude. Even in the conservative Simulation~10, which assumed only 1\% of the seismic energy of the nominal case, the minimum acceleration-to-gravity ratio still reached 0.2 (Fig.~\ref{fig:amax1}d, h). This result suggests that even impacts smaller than the Kamoi event could have triggered global seismic shaking and surface renewal.

Assuming constant seismic efficiency, density, and impact velocity, we estimate that the minimum impactor mass required to induce global seismic shaking on Itokawa is as small as 4.3 g (Fig. \ref{fig:amax1}d, h). According to the standard interplanetary dust flux model \citep{Grun1985}, such collisions are expected to occur approximately once every 20\,000 years for an asteroid the size of Itokawa. Given the estimated space weathering timescale of $\sim$1\,000 years for S-type asteroids, we infer that more than 5\% of sub-kilometer-sized asteroids could statistically exhibit fresh surfaces as a result of a single impact \citep{Jin2022}. This probability is non-negligible. Therefore, it is reasonable to expect that future missions will encounter additional small asteroids with surface characteristics similar to those of Itokawa.

%For larger S-type asteroids, the likelihood of such resurfacing events is even higher.

We note, however, that some previous studies have suggested a much longer space weathering timescale, up to $10^7$ years. If that were the case, even small impacts occurring every $\sim 10^4$ years would suffice to refresh surfaces over geological timescales, and widespread fresh terrains would be common among S-type near-Earth asteroids. However, such features are rare except for Itokawa. This discrepancy may support the notion that space weathering proceeds rapidly, on the order of $\sim 10^3$ years, consistent with our results and sample-based estimates \citep{Nagao2011, Noguchi2014, Jin2022}.

These findings not only explain the exceptional surface freshness of Itokawa, but also underscore the potential of space weathering analysis as a diagnostic tool for investigating both surface and internal properties of asteroids. Our methodology could be extended to other mission-targeted asteroids, such as Dimorphos, the secondary of (65803) Didymos, which was targeted by the DART mission. The kinetic energy of the DART impactor was approximately 2.8 times greater than that of the Kamoi impactor, and Dimorphos's diameter ($\sim$150~m) is about half that of Itokawa \citep{2023Natur.616..461G}. These characteristics suggest that the DART impact may have triggered global seismic shaking on Dimorphos, potentially leading to surface disturbances observable by the Hera mission.

Moreover, future missions such as Hayabusa2's extended mission to (98943) Torifune and OSIRIS-APEX's exploration of (99942) Apophis could benefit from applying our approach, particularly if fresh craters similar to Kamoi are found. Expanding this analysis to multiple rubble-pile bodies will deepen our understanding of the internal diversity and evolutionary paths of small asteroids.

\subsection{Seismic diffusivity ($K$) and the rubble-pile structure of Itokawa}
\label{subsec:K_eta}

In this study, we estimated the seismic diffusivity ($K$) of Itokawa to be in the range of 1\,000–-2\,000~$\mathrm{m^2\,s^{-1}}$. This value was derived from the geographic distribution of landslide features reproduced in our simulations, as shown in Fig.~\ref{fig:x1}. The $K$ value for Itokawa is only about 2\% of those estimated for (433) Eros ($\sim$17 km) and (2867) Steins ($\sim$5 km) \citep{Richardson2020}, suggesting that Itokawa has a more scattering-dominated interior structure than these larger S-type asteroids. This difference is likely attributed to its smaller size and higher porosity. 

Assuming a seismic wave velocity of 100 $\mathrm{m\,s^{-1}}$, based on the impact experiments of \citet{Yasui2015}, the corresponding mean free path is estimated to be 30-–60 m. This implies that the interior of Itokawa likely hosts many tens-of-meter–sized blocks, comparable to the largest surface boulder, Yoshinodai ($\sim$50 $\times$ 30 $\times$ 20 m). This mean free path is significantly shorter than the overall dimensions of Itokawa (535$\times$294$\times$209 m), indicating that seismic waves are strongly scattered at tens-of-meters scales. Itokawa was the first asteroid confirmed to have a rubble-pile structure, based on two lines of definitive evidence: the abundance of surface boulders and high porosity of $\sim$40\% \citep{Fujiwara2006}. However, the internal distribution and size of subsurface boulders remain poorly understood.

Our seismic modeling suggests that the interior is filled with numerous boulders, an order of magnitude smaller than the body size. Assuming a porosity of 40\%, the number of such blocks may range between $10^4$ and $10^5$. Therefore, our study provides unique dynamical evidence based on seismic wave propagation modeling that supports the hypothesis that Itokawa’s interior is truly a rubble pile in nature.

We also estimated the seismic efficiency ($\eta$) for Itokawa to range from $5.0 \times 10^{-8}$ to $5.0 \times 10^{-7}$. This is an order of magnitude lower than the values reported for Eros and Steins \citep{Richardson2020}, and significantly lower than those for the Moon and Mars \citep{Latham1970, Lognonne2020}. We interpret this low efficiency as a consequence of Itokawa’s high porosity and the armoring effect of its boulder-rich surface. In fact, laboratory experiments using unarmored sand targets have yielded higher efficiencies on the order of $10^{-5}$ \citep{Yasui2015, Matsue2020}, which further supports our interpretation of the low $\eta$ value for Itokawa.

\section{Summary}
In this study, we simulated impact-induced seismic shaking and subsequent landslides on asteroid Itokawa to assess the influence of the Kamoi impact on surface space weathering. Our results show that an impactor as small as 4.3 g can induce surface accelerations sufficient to destabilize boulders across the entire surface. The simulations reveal that fresh material originally buried beneath boulders can be exposed through ejecta launch or downslope motion. These findings are consistent with spatial variations in space weathering observed by the Hayabusa spacecraft.

By reproducing the observed surface patterns, we constrained the seismic diffusivity to 1\,000--2\,000~$\mathrm{m^2\,s^{-1}}$, suggesting the presence of interior blocks on the order of 10 meters in size. We also estimated the seismic efficiency to be on the order of $10^{-7}$. The agreement between our estimates and previous studies supports the potential of using space weathering distributions as diagnostic tools for inferring both surface and internal physical properties of asteroids.

\begin{acknowledgements}
This research was supported
by a National Research Foundation of Korea (NRF) grant funded by the Korean government (MEST) (No. 2023R1A2C1006180). SJ was supported by the Research Scholarship for Ph.D. studies from the National Research Foundation of Korea (NRF) grant funded by the Korean Government (RS-2024-00407585).
\end{acknowledgements}

\bibliographystyle{aa}
\bibliography{ref_ito}

\end{document}